\documentclass[12pt]{iopart}
\usepackage[utf8]{inputenc}
\usepackage{graphicx}
\usepackage{amssymb}
\usepackage{hyperref}
\usepackage{datetime}
\usepackage{siunitx}
\usepackage{braket}
\usepackage{cleveref}
\usepackage{gensymb}
\usepackage{xcolor}

\begin{document}

\title[]{Cavity-assisted preparation and detection of a unitary Fermi gas}

\author{K. Roux$^1$, V. Helson$^1$, H. Konishi$^1$ and J.P. Brantut$^1$}

\address{$^1$ Institute of Physics, EPFL, 1015 Lausanne, Switzerland}
\ead{jean-philippe.brantut@epfl.ch}
\vspace{10pt}
\begin{indented}
\item[]
\end{indented}

\begin{abstract}

We report on the fast production and  weakly destructive detection of a Fermi gas with tunable interactions in a high finesse cavity. The cavity is used both with far off-resonant light to create a deep optical dipole trap, and with near-resonant light to reach the strong light-matter coupling regime. The cavity-based dipole trap allows for an efficient capture of laser-cooled atoms, and the use of a lattice-cancellation scheme makes it possible to perform efficient intra-cavity evaporative cooling. After transfer in a crossed optical dipole trap, we produce deeply degenerate unitary Fermi gases with up to $7 \cdot 10^5$ atoms inside the cavity, with an overall $2.85$ \si{\second} long sequence. The cavity is then probed with near-resonant light to perform five hundred-times repeated, dispersive measurements of the population of individual clouds, allowing for weakly destructive observations of slow atom-number variations over a single sample. This platform will make possible the real-time observation of transport and dynamics as well as the study of driven-dissipative, strongly correlated quantum matter.
\end{abstract}

\vspace{2pc}

\vspace{2pc}

\maketitle

\section{Introduction}

A low-temperature gas of interacting Fermions is one of the most complex quantum systems. In spite of its conceptual simplicity, the interplay between the Pauli principle, geometry and interparticle interactions yields very rich structures that underlie most aspects of matter, from nuclei to solid materials and dilute ultra-cold gases. Even though the Hamiltonians describing such systems are often very simple, predicting precisely their properties remains an outstanding challenge. For this reason, the interacting Fermi gas is a natural candidate for quantum simulation, where a well-controlled system is tuned to emulate the Hamiltonian,  and the corresponding properties are read-out by direct measurements, complementing or circumventing the need for numerical or analytical solutions \cite{cirac_goals_2012}.

The last decade has seen spectacular progress in the preparation and detection of ultracold Fermi gases, and in their use for the quantum simulation of complex condensed matter or even high energy physics problems \cite{Esslinger:2010aa,bloch_quantum_2012}. Detection schemes in particular have progressed with the development of quantum gas microscopes \cite{gross_quantum_2017}, which provide high-resolution images of the atomic distribution in a lattice gas. However, all methods so far applied to Fermionic quantum matter are intrinsically destructive as they prevent the acquisition of time-resolved information over a single realization but rather produce snapshots at a given point in time. Measurements of the dynamics of atomic systems therefore rely on sample-to-sample comparisons. Transport experiments, in particular, are limited by the fidelity of sample reproduction \cite{valtolina_josephson_2015,hausler_scanning_2017}, calling for new methods for the continuous, weakly destructive monitoring of individual samples \cite{uchino_universal_2018}.

Cavity quantum-electrodynamics (cQED) offers the best understood platform for studying, performing and controlling weakly or non-destructive measurements \cite{haroche_exploring_nodate}. Indeed, a high-finesse cavity allows for the coherent enhancement of measurement signals over incoherent heating and scattering mechanisms, yielding an increase of the signal-to-noise ratio at fixed destructivity \cite{lye_nondestructive_2003,Hope:2004aa}. When the motional degrees of freedom of atoms can be neglected, a cavity with a cooperativity exceeding unity can be used to realize quantum non-demolition (QND) measurements \cite{haroche_exploring_nodate}. For many-body systems at low energy, such as degenerate, interacting quantum gases, the interplay of measurement back-action with strong interactions has not been explored experimentally beyond the effects of spontaneous emission \cite{Schneider2017,Tomitae2017,bouganne_anomalous_2020}, in spite of strong theoretical interest \cite{Eckert:2008aa,Roscilde:2009aa,Bernier2013,Daley:2014ab,ashida_quantum_2016,Mazzucchi:2016ab,Mazzucchi:2016aa,Ashida:2018ab,Sorensen:2018aa,uchino_universal_2018}. High-finesse cavities have been successfully combined with evaporative cooling, allowing for the coupling of weakly interacting ultra-cold gases, Bose-Einstein condensates \cite{colombe_strong_2007,brennecke_cavity_2007,gupta_cavity_2007,slama_superradiant_2007}, lattice Bose gases \cite{landig_quantum_2016,klinder_observation_2015} to few-photon fields, and more recently ultra-cold Fermi gases \cite{roux_strongly_2020}. 

In this paper, we present a  novel experimental setup which combines a high-finesse cavity and a strongly interacting Fermi gas of $^6$Li. In particular we present how the same cavity is used both at the preparation and detection stages. For preparation, the cavity is integrated in a compact, single-chamber vacuum setup and combined with standard laser cooling methods, thus limiting the technical cost compared with other lithium machines without cavities. The number of atoms prepared and the cycle time achieved in our system are competitive with state-of-the-art apparatus making use of advanced laser cooling methods \cite{Burchianti:2014ab,Long:2018ab}. To illustrate the potential of the cavity as a detection tool for Fermi gases, we demonstrate repeated, weakly destructive measurements of the time evolution of the number of atoms in an individual cloud. This represents a key milestone towards quantum-limited transport measurements, for example in a two-terminal configurations \cite{Krinner:2017aa,uchino_universal_2018}.

The paper is structured as follows: in section 2, we describe the vacuum and laser systems, as well as the different electromagnets accommodated on the apparatus in order to produce high magnetic fields. In section 3 we present the parameters of the  high finesse cavity at different wavelength. In addition we detail our design that allows to combine the high finesse cavity together with large magnetic fields as well as the cavity length stabilization technique. In section 4, we demonstrate an efficient cooling strategy, using a cavity-enhanced dipole trap with a lattice cancellation scheme combined with a crossed optical dipole trap. Our procedure produces a unitary superfluid within the cavity mode with up to $7\cdot 10^5$ atoms at $T<0.1\,T_F$, with $T$ and $T_F$ its temperature and Fermi temperature. In section 5 we measure the atom-induced dispersive shift of the cavity resonance frequency, and demonstrate the low heating and atomic losses induced by five hundred-times repeated measurements over an individual Fermi gas sample. Last, section 6 presents some of the perspectives opened by this system regarding quantum simulation with ultracold atoms.
\section{Overview of the apparatus}

\begin{figure*}[ht]
\begin{center}
  \includegraphics[width=0.8\textwidth]{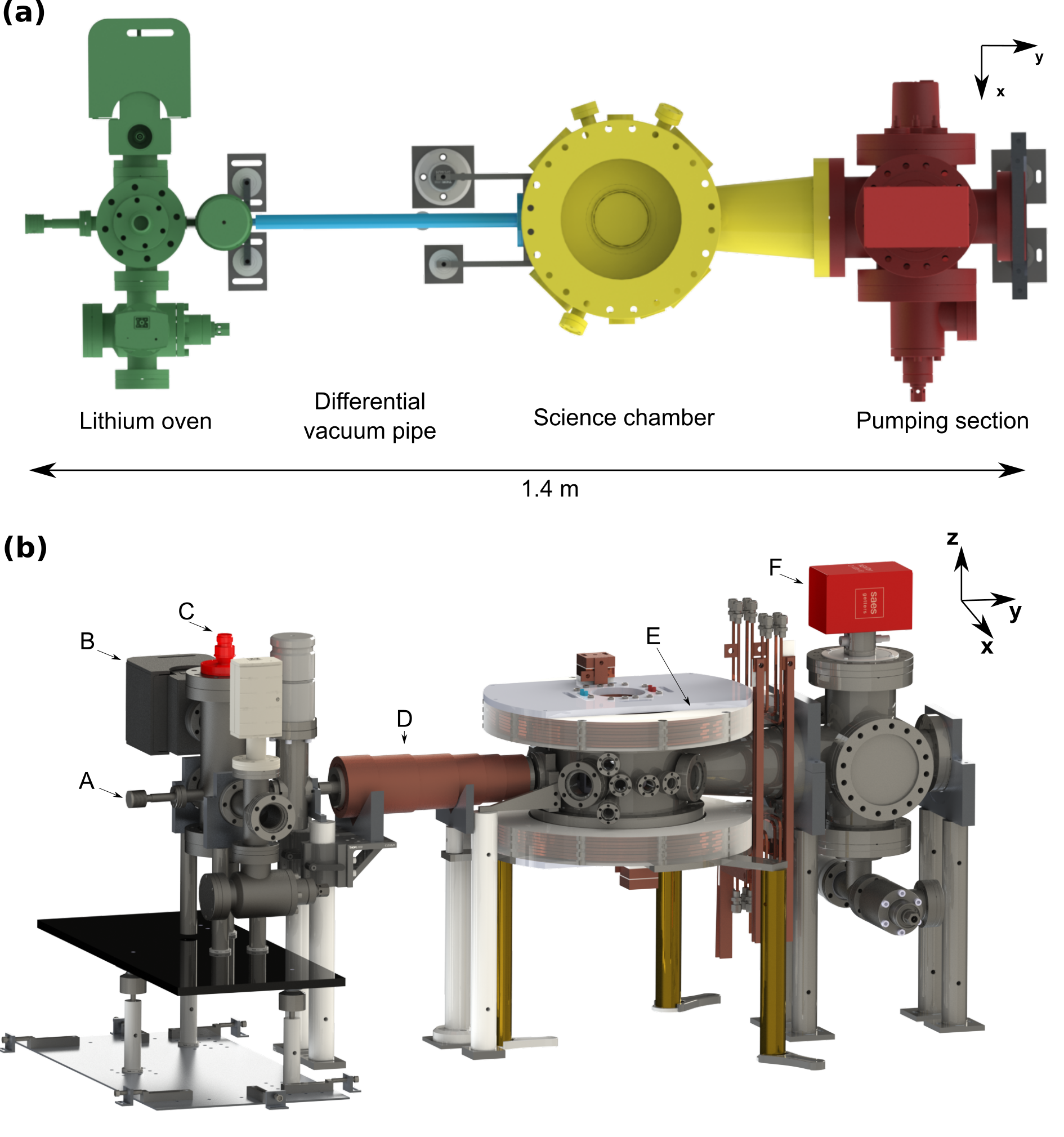}
    \caption{\textbf{An apparatus for cavity QED and strongly interacting Fermi gas.}
       \textbf{(a)} Top-view representation of the vacuum system. Colors refer to different functions of the apparatus: oven section (green), conical pipe ensuring differential pumping (blue), science chamber (yellow) and pumping section (red). 
   \textbf{(b)} Overview of the apparatus. On the oven section where the lithium oven (A) is placed, high vacuum is maintained by an ion pump (B, \textit{Agilent VacIon 20})  and a getter pump (C, \textit{SAES Capacitor D400}). A slow beam of atoms is created by a Zeeman slower (D). The science chamber features an ensemble of electromagnets for magneto-optical trapping and for the use of Feshbach resonances (E). Ultra-high vacuum in this section is ensured by a combined getter and ion pump (F,\textit{SAES NexTORR D1000-10}).
    }
    \label{fig:setup}
\end{center}
\end{figure*}

The apparatus is based on a single science chamber maintained in ultra-high vacuum where all cooling and measurement procedures are performed. As depicted in figure \textbf{\ref{fig:setup}(a)} and \textbf{\ref{fig:setup}(b)} the vacuum system comprises four different sections. The science chamber itself is based on a cylinder with CF200 flanges with two reentrant viewports on top and bottom, hosting electromagnets and offering a numerical aperture up to $0.65$. In addition to connections with the other sections, it has four CF40 viewports and ten CF16 viewports which offer plenty of optical access for laser cooling, dipole trapping and cavity beams. The oven section is separated from the science chamber by a gate valve and a $40$ \si{\centi\meter}-long conical pipe to ensure a good differential vacuum. The science chamber is connected via a conical extension to a pumping section featuring a combined ion and getter pump (SAES Getters) and electrical feedthroughs. To assess the quality of the vacuum, we measured a vacuum-limited lifetime of laser cooled clouds of $40$ \si{\second}.

\begin{figure*}[h!]
\begin{center}
  \includegraphics[width=0.6\textwidth]{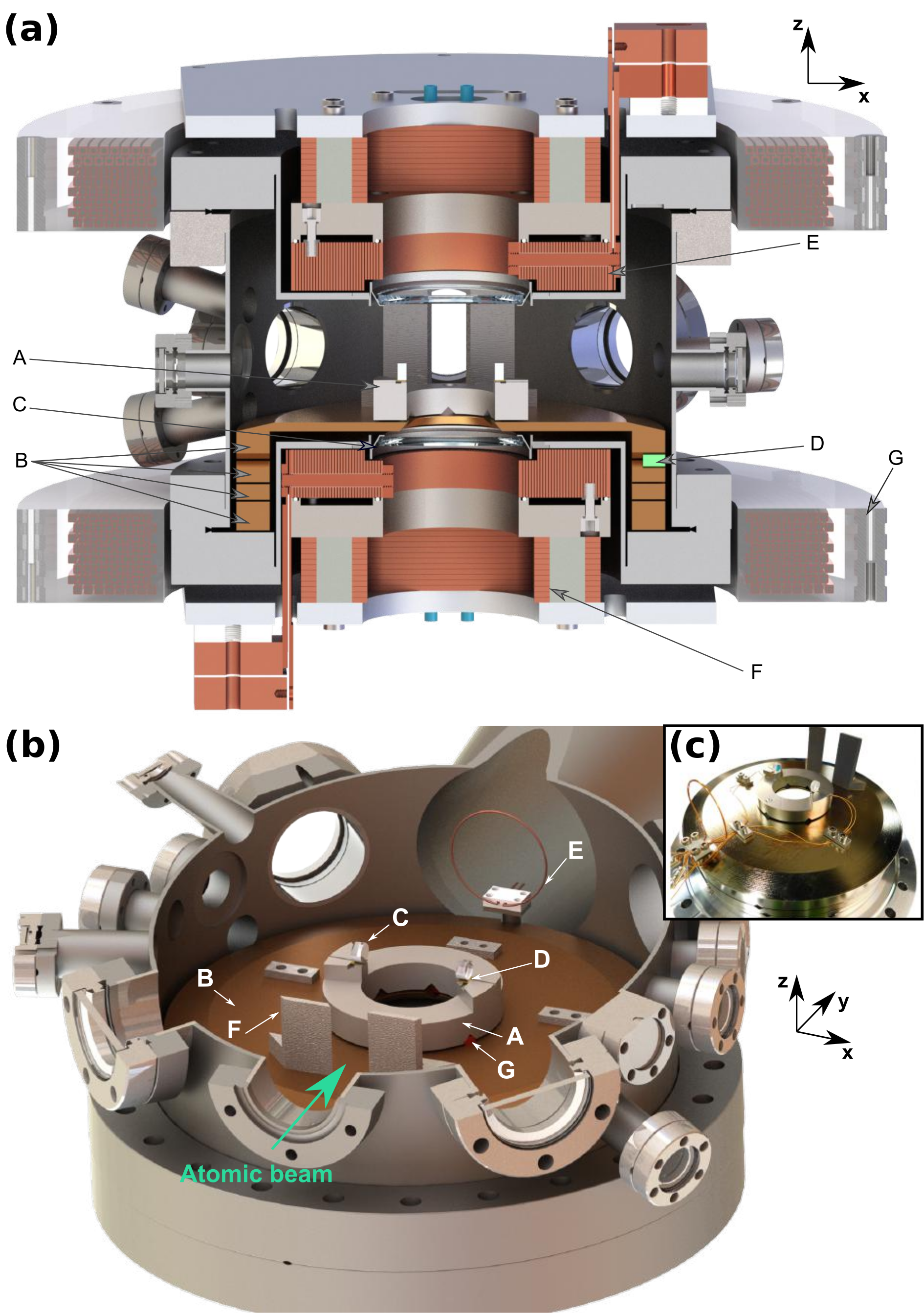}
    \caption{\textbf{Detailed view of the science chamber.}
       \textbf{(a)} Vertical cut view of the science chamber. The science platform (A) hosting the cavity mirrors lies on several stages of vibration isolation made of non-magnetic stainless steel (B) separated by Viton dampers (D). On the top and bottom, reentrant viewports (C) allow for a large optical access, up to a numerical aperture of 0.63. An ensemble of two pairs of coils (G) provides a magnetic field gradient for magneto-optical trapping and a homogeneous bias. In the reentrant viewport a compact ensemble of electromagnets combines compact bulk-machined coils (E), allowing to reach up to $1200$ G with a finite curvature, and cloverleaf coils (F) to precisely align the magnetic field saddle point on the cavity mode position.
       \textbf{(b)} View of the science chamber with the cavity mirrors (C) and piezoelectric actuators (D) glued on a stiff titanium holder (A). The holder itself is isolated from the isolation stacks (B) via Viton dampers (G). A single loop radio-frequency antenna (E) is positioned close to the cavity. Titanium shields (F) avoid direct coating of the mirror by the beam of atoms coming from the oven.  
       \textbf{(c)} Photograph of the science cavity ensemble prior to installation in the vacuum system.
    }
    \label{fig:cavity}
\end{center}
\end{figure*}

The science chamber is equipped with two main sets of electromagnets that can be seen in figure \ref{fig:cavity}. The first set consists of two pairs of large hollow-copper wound coils, positioned outside of the chamber, in the Helmholtz configuration. One of these pairs creates the magnetic quadrupole of the magneto-optical trap (MOT). The field of these coils connects adiabatically to the decreasing bias of the Zeeman slower coil, ensuring the shortest possible distance between the Zeeman-slowed atomic beam and the MOT. The second pair is used to create a homogeneous magnetic field bias up to $250$ G with a curvature $< 1$ \si{\hertz}. The second set of electromagnets, located inside the reentrant viewports, mainly consists of a pair of custom-built, bulk-machined coils \cite{roux_compact_2019}, designed to exploit the broad Feshbach resonance of $^6$Li at $832\,$G. Their configuration departs from Helmholtz, thereby creating a magnetic saddle point with a negative curvature along the horizontal direction, as commonly encountered in ultracold lithium experiments (see for example \cite{ketterle_making_2008}). On top of these coils, four pairs of compensation coils in the cloverleaf configuration are used to position the magnetic saddle point in the horizontal plane.

The near-resonant light at $671$ \si{\nano\meter} for laser cooling and imaging is derived from diode lasers. A tapered amplified diode (Toptica TA Pro) is frequency-stabilized onto the D2 transition of $^6$Li using saturated absorption spectroscopy on a home-built lithium vapour cell. This laser provides all the laser cooling light for the MOT cooling stage. Another tapered amplifier is offset-locked onto the first and provides light for the Zeeman slower. A separate diode laser (Toptica DL Pro) detuned by several hundreds of \si{\mega\hertz} is used to perform absorption imaging at high magnetic fields. The dipole traps are derived from an ultra-narrow, frequency tunable fiber laser at $1064$ \si{\nano\meter} (NKT Photonics Adjustik), amplified to $8.1$ W using an Yb-doped fiber amplifier (Azurlight ALS-IR-1064-20-A). A small portion of the light is injected into a fiber-pigtailed frequency doubler (NTT electronics WH-0532), generating light at $532$ \si{\nano\meter}. This light and part of the $1064$ \si{\nano\meter} are routed to the high-finesse cavity to stabilize its length and generate an intra-cavity dipole trap. The remaining power ($6$ W) is split evenly to create the two independent arms of the crossed dipole trap. Acousto-optic modulators (AOM) are used to control the beam powers as well as to frequency-shift the two arms of the crossed trap to avoid unsought interference effects. Details and schematics of laser setups can be found in Appendix A.


\section{High-finesse cavity} 

\subsection{Optical properties}

The core of the apparatus is a $4.1$ \si{\centi\meter}-long Fabry-Perot cavity formed by two $6$ \si{\milli\meter} diameter dielectric mirrors (Advanced Thin Films). The optical properties of the cavity are summarized in table \ref{table:properties}. The design of the cavity geometry results from a trade-off between (i) the possibility to accommodate sufficiently large laser cooling beams intersecting at the cavity center thus enforcing a minimal inter-mirror distance, (ii) the use of the cavity to produce an optical dipole trap with a large beam waist to efficiently capture atoms and (iii) the need for a high cooperativity, allowing to work in the strong light-matter coupling regime.

The cavity has a high-finesse at $671$ \si{\nano\meter}, the wavelentgh which addresses the $2S \longrightarrow 2P$ transitions of $^6$Li. It is stabilized using light at $532$ \si{\nano\meter}, where the cavity has a moderate finesse, which ensures that the far detuned stabilization beam weakly affects the atomic density and does not lead to spurious trapping effects. The cavity is also resonant at $1064$ \si{\nano\meter}, where it is used to produce an intra-cavity optical dipole trap (see Section 4.1).

\begin{table}[ht]
\centering
\caption{Optical properties of the cavity.}
\begin{tabular}{ |p{5cm}||c|c|c| }
\hline
&$671 \,$nm &$532 \,$ nm & $1064 \,$ nm \\
\hline
\hline
Cavity length & \multicolumn{3}{|c|}{$4.131(1) \,$ \si{\centi\meter}} \\
\hline
Free spectral range (FSR) & \multicolumn{3}{|c|}{$3.6277(1) \,$ \si{\giga\hertz}} \\
\hline
Mirror diameter & \multicolumn{3}{|c|}{$6 \,$ \si{\milli\meter}} \\
\hline
Finesse & $4.7(1) \cdot 10^{4}$ & $2.4(1) \cdot 10^{3}$ & $3.6(1) \cdot 10^{3}$ \\
\hline
Linewidth $\kappa/2\pi$  & $0.077(1) \, \si{\mega\hertz} $ &  $1.5(1) \,$\si{\mega\hertz} & $1.0(1) \, $ \si{\mega\hertz} \\
\hline
TEM$_{00}$ Mode waist & $45.0(3)$  \si{\micro\meter}  & $40.1(3)$ \si{\micro\meter} & $56.6(3)$ \si{\micro\meter}\\
\hline
\end{tabular}
\label{table:properties}
\end{table}

The atom-photon coupling is characterized by the cavity-QED parameters \cite{tanji-suzuki_chapter_2011} $(g,\kappa,\Gamma)=2\pi\times (0.479,0.077,5.872)$ \si{\mega\hertz}. Here $g$ is half of the vacuum Rabi splitting for a single atom located at the field maximum, for the closed D2 $\sigma_-$ transition $\ket{2S_{1/2},m_{J}=-1/2} \longrightarrow \ket{2P_{3/2},m_{J}=-3/2}$ relevant at high magnetic fields. $\kappa$ is the intra-cavity intensity decay rate and $\Gamma$ the natural linewidth of the $2P$ excited states of $^6$Li. The cavity parameters yield a single-atom single-photon cooperativity $4g^2/\kappa \Gamma=2.02$, reaching the strong coupling regime.

The cooperativity measures the ratio between photon scattering rates into the cavity and in free space, and the strong coupling is achieved when this quantity exceeds unity. However, for photons close to the atomic resonances, the two orders of magnitude difference between the atomic and cavity linewidth yield weak optical signals, as scattering by the atoms into free space dominates over transmission through the mirrors. While this is usually detrimental for quantum information applications \cite{Miller:2005aa,Reiserer:2015aa}, the narrow linewidth of the cavity is expected to provide an advantage for weakly-destructive measurements in the dispersive regime \cite{Yang:2018ab}. With $\kappa < 4 E_\mathrm{r}$, $E_\mathrm{r}$ being the recoil energy, the measurement back-action is suppressed even for strongly interacting gases \cite{uchino_universal_2018}.

For our experimental configuration, two effects reduce the light-matter coupling strength. First, due to the magnetic field orientation being orthogonal to the cavity axis, the $\sigma_-$ transition has to be driven with light linearly polarized in the horizontal plane thereby reducing the coupling by $\sqrt{2}$. Second, the atoms are smoothly distributed along the cavity direction, such that the average over the cosine square-shaped mode function of the cavity field  further reduces the average coupling by a factor of $2$.

\subsection{Science platform}

As shown in figure \textbf{\ref{fig:cavity}}, the science platform hosting the cavity rests inside the science chamber such that the cavity waist is approximately at the geometric center of the chamber. It rests on a four-stages vibration damper, each stage made of heavy non-magnetic stainless steel rings stacked onto Viton dampers \cite{Ottl:2006aa}. The ensemble rests on the side parts of the reentrant viewport (see figure \textbf{\ref{fig:cavity}(a)}). The first three stages enclose the reentrant section of the bottom viewport, and the last ring covers the viewport except for the optically accessible aperture. The optical cavity mirrors are glued (Masterbond EP21TCHT-1) onto shear piezo-electric actuators (Noliac CSAP02-C03), themselves fixed onto the ring-shaped titanium science platform. The cavity is aligned prior to gluing, such that no clamping or adjustable parts are installed with the cavity in the chamber. Shields are fixed on the last isolation stage to protect the cavity mirrors against direct coating from the lithium beam exiting the oven. A radiofrequency antenna, attached to the bottom isolation stack for mechanical isolation, is positioned $\sim10$ \si{\centi\meter} away from the atoms to drive hyperfine transitions. The inter-mirror distance allows for the $2.5$ \si{\centi\meter} diameter MOT beams to cross within the cavity volume, intersecting the cavity axis at an angle of $45$\degree.

\subsection{Cavity laser system}

A dedicated laser system stabilizes the absolute length of the cavity and ensures relative stability between the cavity probe beam, the cavity trap beams and the cavity resonance. The cavity probe beam at $671$ \si{\nano\meter} is derived from a reduced-linewidth laser diode (Toptica DL Pro), narrowed down to $\sim 10$\,kHz by stabilizing its frequency onto a narrow ($30$ \si{\kilo\hertz}), home-built transfer cavity using the Pound-Drever-Hall (PDH) scheme (see Appendix A for details). The beam passes through a wide-band, double pass AOM (AAoptoelectronic MT110-B50A1-VIS) to allow for fast sweeps of the frequency and power of the beam.

The science cavity length is locked onto the $532$ \si{\nano\meter} light, the second harmonic of the $1064$ \si{\nano\meter} laser. To ensure the relative stability of the probe and cavity-trap lasers with respect to the cavity, we first stabilize the transfer cavity onto the $1064$ \si{\nano\meter} laser. To allow for fine-tuning of the lock point, the beam goes through a broadband fibered electro-optic phase modulator (NIR-MPX-LN-05), and a sideband is used as reference for the transfer cavity. We then stabilize the probe laser onto the transfer cavity. A high precision wavemeter (High Finesse WS8-2) referenced to the lithium spectroscopy is used to monitor the frequency of the probe laser at long timescales. It generates a control signal acting on the $1064$ \si{\nano\meter} laser, which thus adapts its frequency so as to maintain the science cavity resonant with the probe laser. 

The stabilization beam of the science cavity at $532 \,$ \si{\nano\meter} is maintained at constant power throughout the entire experimental process, thereby generating a spurious repulsive potential for the atoms. We mitigate this effect using two methods. On one hand, we use a very deep phase modulation for the PDH stabilization, such that the sidebands concentrate most of the laser power compared to the carrier. Once the cavity length is stabilized the sidebands are reflected which limits the intracavity optical power. On the other hand, we use the TEM$_{02}$ mode of the cavity for stabilization, to spread the intracavity power over a larger volume, further minimizing the residual potential depth. This allows for keeping the residual peak dipole potential below $46$ nK.


\section{Fast production of a unitary Fermi gas}

\begin{figure}[ht]
\begin{center}
    \includegraphics{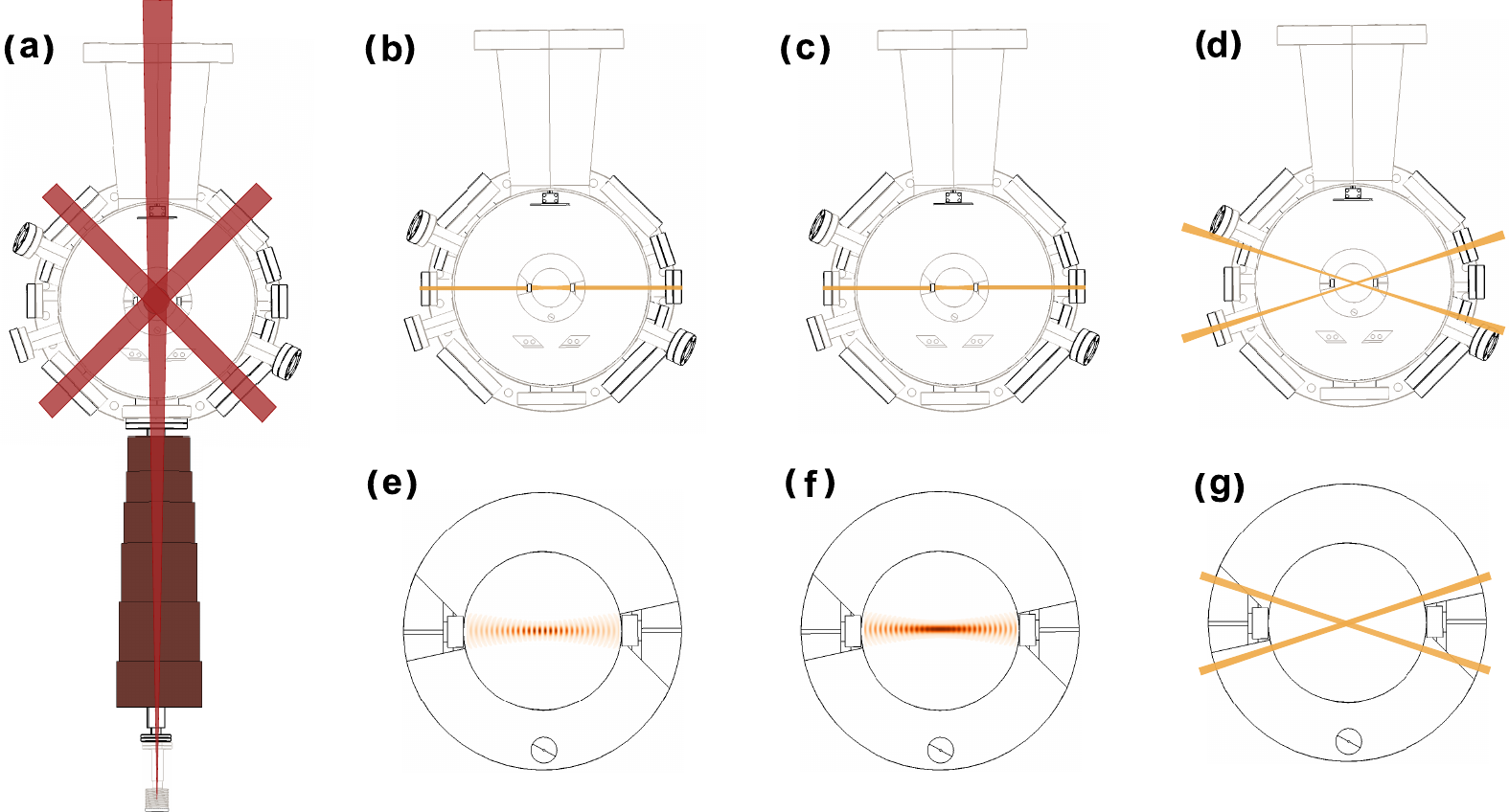}
    \caption[Cooling procedure]{ \textbf{Overview of the cooling cycle.}
    Panels \textbf{(a)-(d)} show a top-view of the chamber while panels \textbf{(e)-(g)} are close top-view of the cavity at different cooling steps. The atoms are slowed down by the Zeeman slower and captured in a MOT \textbf{(a)}. After compression of the MOT, atoms are loaded in an intracavity dipole trap (\textbf{(b)}, \textbf{(e)}) where the first evaporative cooling stage takes place. The sidebands imprinted by an electro-optic modulator onto the cavity dipole trap beam suppress the lattice structure of the trap in the cavity direction (\textbf{(c)}, \textbf{(f)}) enhancing the efficiency of the evaporation. Atoms are then transferred into the crossed dipole trap (\textbf{(d)}, \textbf{(f)}), where the last evaporation is performed to reach quantum degeneracy.}
    \label{fig:cooling}
\end{center}
\end{figure}

The cooling sequence is outlined in figure \textbf{\ref{fig:cooling}}. We start with atoms slowed down by the Zeeman slower and then trapped in a MOT realized on the D2 line of $^6$Li  (figure \textbf{\ref{fig:cooling}(a)}). After $1.3$ \si{\second} of loading, a $50$ \si{\milli\second} stage of compression is applied by ramping the MOT beams frequencies close to the atomic resonance and by reducing their power to about $5\%$ of their initial value. This creates a dark and dense cloud close to the Doppler temperature which is then loaded in an intra-cavity optical dipole trap. 

\subsection{Cavity dipole trap}

\begin{figure*}[h]
\begin{center}
    \includegraphics{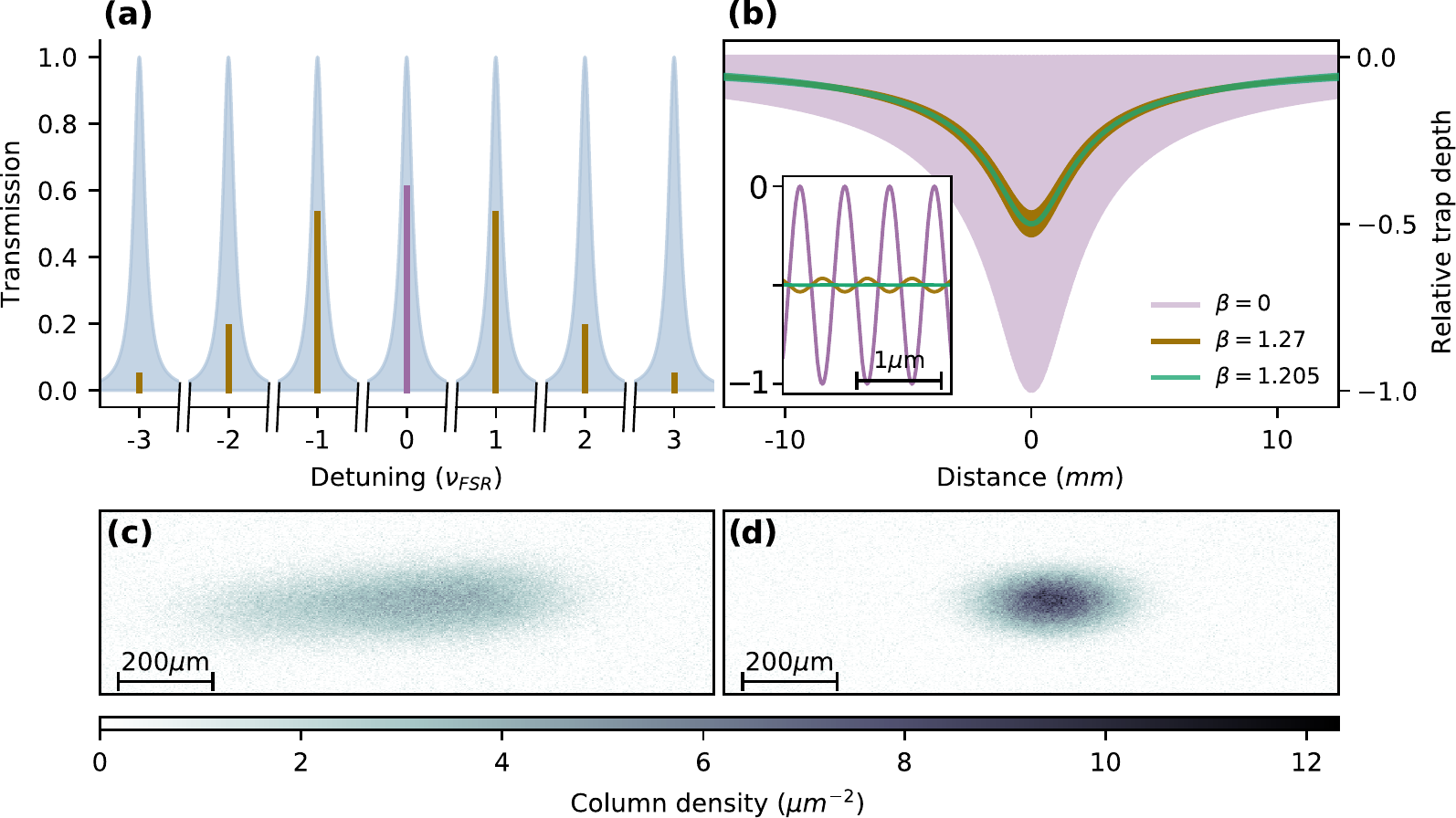}
    \caption[Lattice suppression]{\textbf{Suppression of the lattice structure in a cavity dipole trap.}   
\textbf{(a)} Sidebands are imprinted on the cavity dipole trap beam using phase modulation at a frequency corresponding to the cavity free spectral range with an optimum modulation depth set to $1.27$ rad. \textbf{(b)} View of the dipole potential created by the cavity dipole trap in the cavity. The dipole potential without the phase modulation exhibits a strong lattice structure (pink curve). The potential resulting from the combination of the different sidebands leads to a flat potential at the center. Inset: zoom on the center of the cavity. The green curve shows the optimal theoretical modulation depth of $1.205$ rad while the brown curve ($\beta = 1.27$ rad) shows the optimal modulation used on the experiment which maximize the phase-space density after intracavity evaporation. \textbf{(c)} and \textbf{(d)} Absorption images of unitary Fermi gases with a $2$ \si{\milli\second} of time-of-flight, after evaporative cooling respectively without and with phase modulation of the cavity dipole trap. The images along the horizontal plane with an angle of $11$\degree $\,$  with respect to the cavity axis. The atom number increases by a factor $2$ and the longitudinal size of the cloud reduces by a factor $3$ in the presence of the modulation as shown in \textbf{(d)}.
    }
    \label{fig:sidebands}
\end{center}
\end{figure*}

All-optical cooling of $^6$Li to quantum degeneracy \cite{granade_all-optical_2002} requires efficient loading in an optical dipole trap from the laser cooling stage. This can be efficiently achieved by improving laser cooling using grey molasses \cite{grier_ensuremathlambda-enhanced_2013,burchianti_all-optical_2015} and through the use of very high power lasers \cite{granade_all-optical_2002}. The power build-up offered by cavities has proven to be an elegant alternative \cite{mosk_resonator-enhanced_2001}, avoiding the need for very high intensity lasers while in addition guaranteeing a clean and reproducible beam profile. 

The resonance of the cavity at $1064$ \si{\nano\meter}, where the finesse is $3.6 \cdot 10^3$ (see Table \textbf{\ref{table:properties}}) yielding a circulating power of $132$ \si{\watt} for $244$ \si{\milli\watt} of incident power, realizes this function in the setup. We use the mode structure of the cavity to mitigate the main drawbacks of cavity dipole traps, namely the fixed trap volume and the lattice structure. In order to increase the trap volume, we inject the cavity on the TEM$_{01}$ mode \cite{Naik:2018aa}, with a nodal line oriented in the vertical direction. In addition to increasing the trap volume and thus optimizing the overlap with the MOT, this also decreases the peak laser intensity on the mirror surface. Indeed we observe that for incident powers exceeding $\sim100$ \si{\milli\watt} on the fundamental mode, the cavity displays strong thermal non-linearities, yielding a bi-stable behavior, destabilizing the cavity locking system. We found that the use of the TEM$_{01}$ mode improves the stability (see Appendix for details). The cavity dipole trap is on during the MOT loading and compression. After switching off the laser cooling beams, we typically obtain $1 \cdot 10^7$ atoms at $\sim1$ mK.

As cavity dipole traps present a lattice structure along the cavity axis due to the standing wave nature of the cavity modes, atoms cannot redistribute between sites and accumulate in the deepest trap regions thereby crippling the efficiency of evaporative cooling. We circumvent this problem using a lattice cancellation scheme previously demonstrated for the uniform interrogation of thermal atoms held in cavities \cite{cox_spatially_2016,Vallet:2017aa}. At the center of a symmetric cavity, two consecutive longitudinal modes form standing waves with a relative dephasing of $\pi/2$, so that the sum of intensities is effectively homogeneous. We use a free-space electro-optic phase modulator (QUBIG PM9-NIR) driven at the frequency of the longitudinal mode spacing, to create sidebands driving the neighboring longitudinal modes in addition to the carrier, as depicted in figure \textbf{\ref{fig:sidebands}(a)}. The large frequency difference between consecutive sidebands ensures that the dipole potential is the sum of the contribution of each components. With $33\%$ of the total power equally distributed in the sidebands, the total dipole potential is effectively flat at the location of the atoms, as shown schematically in figure \textbf{\ref{fig:sidebands}(b)}.

In practice, we first load the atoms from the MOT into the lattice trap (figure \textbf{\ref{fig:cooling}(b))} which maximizes the loading efficiency, and then the magnetic field is ramped to the Feshbach resonance of the two lowest hyperfine states denoted $\ket{\uparrow}$ and $\ket{\downarrow}$ respectively. After the first evaporation ramp of $150$ \si{\milli\second} bringing the cavity circulating power to $13.3$ \si{\watt}, we transfer the atoms into the lattice-free trap (figure \ref{fig:cooling}(c)), derived from the same laser and incident on the cavity with crossed polarization. Then the second $150$ \si{\milli\second} ramp of evaporative cooling is performed where the circulating optical power of the cavity trap is reduced to $960$ \si{\milli\watt}. 

The figures \textbf{\ref{fig:sidebands}(c)} and \textbf{(d)} illustrate the improvement offered by the lattice cancellation scheme compared with the same procedure without the cancellation scheme. The absorption images are taken after a $2$ \si{\milli\second}  time-of-flight. The redistribution of atoms along the cavity axis yields a reduction of the longitudinal cloud size by a factor of $3$ without atom losses. While the size cannot be directly related to temperatures in the unitary regime, it reflects the release energy which is monotonically related to temperature in a trapped gas \cite{OHara:2002aa,:2003aa,Stewart:2010aa}. The size reduction and increase in density visible in the figure indicates a significant improvement of the evaporative cooling efficiency. In the next stage of the experimental cycle, we combine this broad cavity trap with a crossed dipole trap intersecting at the cavity mode position. We observed that the lattice cancellation scheme increases the number of atoms transferred to the crossed dipole trap by a factor of $2$ with all other parameter being identical.

\subsection{Quantum degenerate, unitary Fermi gas}

\begin{figure*}[ht]
\begin{center}
    \includegraphics{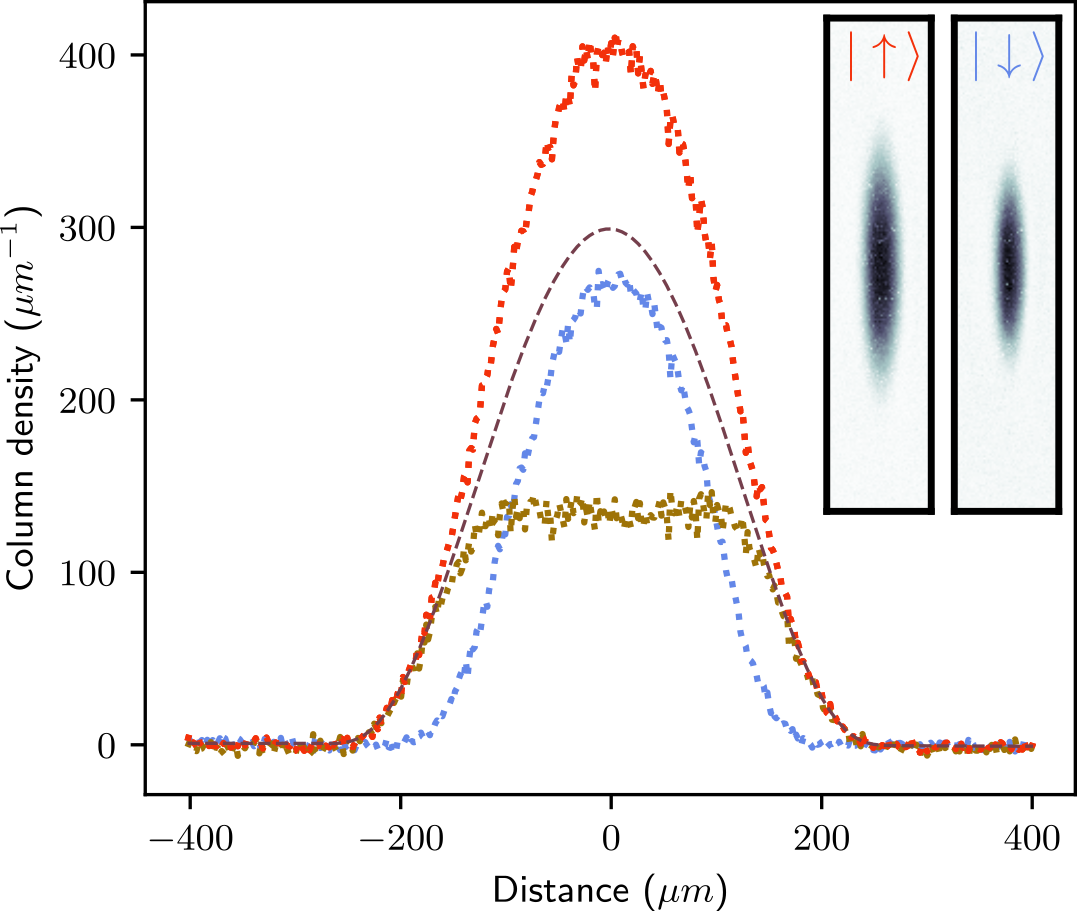}
    \caption[temperature]{\textbf{Superfluidity and temperature measurement using a spin-imbalanced Fermi gas.}
    Column density of both spin components after a short time-of-flight of $0.6$ \si{\milli\second}, with $1.1 \cdot 10^5$ and $7.5 \cdot 10^4$ atoms in $\ket{\uparrow}$ (red) and $\ket{\downarrow}$ (blue), respectively. The density difference (brown) exhibits a  plateau characteristic of a paired superfluid core surrounded by a polarized shell. The fully polarized wings of the majority component is fitted with a Thomas-Fermi profile for a non-interacting Fermi gas (purple dashed) yielding a temperature of $0.07(1) \, T_F$. Inset: 2D density distribution of the two spin components after a short time of flight of $0.6$ \si{\milli\second}.
    }
    \label{fig:temperature}
\end{center}
\end{figure*}
 
During the evaporative cooling procedure we combine the lattice-free cavity dipole trap and the crossed optical dipole trap, as shown in figure \textbf{\ref{fig:cooling}(c)} and \textbf{(d)}. Each of the crossed dipole trap beam is focused on the cavity mode position with a waist of $33$ \si{\micro\meter}, and intersect with an angle of $22$\degree $\,$ in the horizontal plane, creating a trap elongated along the cavity direction. During the second intracavity evaporative cooling stage atoms are smoothly transferred to the cross dipole trap with $1$ \si{\watt} of optical power in each arm. Even though the cavity dipole trap produces two separated clouds held in the two lobes of the TEM$_{01}$ mode, both clouds are efficiently collected in the crossed dipole trap after intracavity evaporation is ended.

Once all atoms are transferred in the crossed dipole trap, the last $350$ \si{\milli\second} evaporation ramp is performed to reach quantum degeneracy by reducing the power down to $\sim 11$ \si{\milli\watt} in each arm. The entire evaporation procedure lasts for $800$ \si{\milli\second}, and typically prepares $3.5 \cdot 10^5$ atoms per spin component in a spin-balanced mixture, for a total sequence time of $2.85$ \si{\second}.

For applications in cavity QED, the cloud size needs to be much smaller than the cavity waist, which is achieved in our crossed dipole trap. This however prevents us from using in-situ imaging for thermometry because the optical density of the cloud is too large. For diagnostics of the degenerate gas, we use a reshaped trap formed by a single arm of the crossed dipole trap and the curvature of the magnetic field, with frequencies $2\pi \times (28.5(1),441(7),441(7))$ \si{\hertz}. We transfer the cloud by adiabatically turning off one of the arm within $300$ \si{\milli\second} after the evaporation. 

To measure the temperature of the cloud and observe superfluidity, we use a spin imbalanced gas produced by introducing controlled losses in state $\ket{\downarrow}$ using the \textit{p}-wave Feshbach resonance at $210$ G prior to the evaporation procedure. After evaporation and transfer in the single-arm trap, we measure the doubly integrated density along the long direction of the cloud, revealing the pressure profile in the trap center \cite{Ho:2009ab,nascimbene_exploring_2010}. A time-of-flight of $600$ \si{\micro\second}, short compared to the longitudinal trap period, is left before imaging to reduce the peak optical density before absorption imaging.

A typical image is shown in figure \textbf{\ref{fig:temperature}}, where the two spin components have been imaged in the same conditions. In the central part of the cloud, the line-density difference between the two spin states is constant, signaling the onset of phase separation between a fully paired, superfluid core and partially polarized wings \cite{partridge_pairing_2006,shin_observation_2006}. At unitarity, the phase diagram of the gas has been studied in details, and phase separation is unambiguously associated with superfluidity \cite{Shin:2008aa}. 

The fully polarized wings of the cloud are described by the ideal-gas equation of state, which can thus be fitted to the density profile to provide unbiased thermometry \cite{ketterle_making_2008}. Such a fit is presented as a dashed line in figure \textbf{\ref{fig:temperature}}, yielding a temperature estimate of $0.07(1)\,T_F$, well below the superfluid transition temperature for the balanced gas $T_c = 0.217\,T_F$.


\section{Weakly destructive probe for strongly interacting Fermi gas} 

\subsection{Dispersive coupling}

In the regime where the cavity resonance is far-detuned from the atomic one, the cavity field couples dispersively to the atoms yielding a shift of the cavity resonance frequency given by $\delta = \tilde{N}g_{0}^{2}/\Delta_a$ \cite{tanji-suzuki_chapter_2011} with $\tilde{N}$ the effective number of atoms coupled to the cavity mode, $g_0$ the single-photon single-atom coupling strength and $\Delta_a$ the detuning between the cavity resonance and the atomic transition. When the center-of-mass motion of the atoms does not enter the dynamics, such as with high temperature atoms or tightly confined atoms, this regime allows for a QND measurement of atom number. In the low temperature regime, a fortiori for our quantum degenerate gases, the recoil associated with measurement back-action breaks the QND character of the measurement, which can nevertheless approach the non destructive regime in the narrow cavity limit \cite{Yang:2018ab,uchino_universal_2018}. 

In the experiment, $\tilde{N}$ differs from the total atom number due to the reduced overlap between the cavity mode and the atomic cloud. The primary contribution to this reduction is a factor $0.5$ due to the averaging of the standing wave along the cavity axis. Further reduction by a factor $\sim0.95$, depending on the trap configuration and atom number, arises due to the transverse extension of the cloud. 

We use light linearly polarized along the magnetic field direction, which couples to the $\pi$-transition, for dispersive measurements. To calculate the dispersive shift we account for the D2 and D1 transitions $\ket{2S_{1/2},m_J = -1/2} \longrightarrow \ket{2P_{3/2},m_J = -1/2}$ and $\ket{2S_{1/2},m_J = -1/2} \longrightarrow \ket{2P_{1/2},m_J = -1/2}$, and their respective detunings $\Delta_{D2\pi}$ and $\Delta_{D1\pi}$ and light matter coupling strengths $g_{D2\pi}$ and $g_{D1\pi}$, such that 
\begin{equation}
\delta = \tilde{N} \left( \frac{g^2_{D1\pi}}{\Delta_{D1\pi}} + \frac{g^2_{D2\pi}}{\Delta_{D2\pi}}  \right).
\end{equation}
At a magnetic field of $832$ G, we have $g_{D1\pi} = 0.576 \cdot g_0$ and $g_{D2\pi} = 0.816 \cdot g_0$ with $g_0 = 2 \pi \times 479$ \si{\kilo\hertz} the coupling strength for a single atom located at the field maximum, for the closed D2 $\sigma_-$ transition $\ket{2S_{1/2},m_{J}=-1/2} \longrightarrow \ket{2P_{3/2},m_{J}=-3/2}$.

\begin{figure*}[ht]
\begin{center}
    \includegraphics{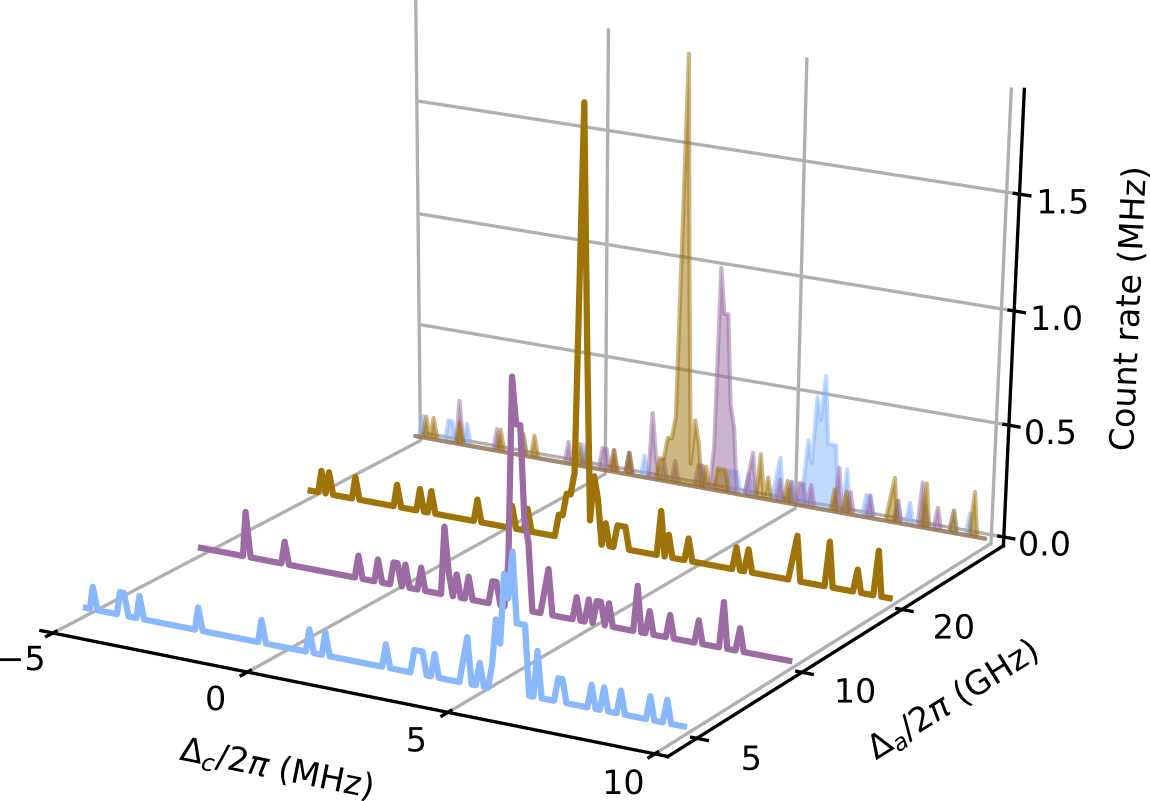}
    \caption[Dispersive shifts]{\textbf{Probing dispersively a unitary Fermi gas.}
    Dispersive measurements for an imbalanced Fermi gas comprising $2.2 \cdot 10^5$ and $1.35 \cdot 10^5$ spin up and down atoms. $\Delta_c$ represents the detuning between the probe and the empty cavity resonance. Raw histograms of transmitted photon numbers with a probe swept over $20$ \si{\mega\hertz} around the bare cavity resonance within $2$ \si{\milli\second} are presented, for three different detunings from the atomic resonance  $\Delta_{D2\pi} = 6.55$ (blue), $11.55$ (purple) and $21.55$ \si{\giga\hertz} (brown).   }
    \label{fig:shift}
\end{center}
\end{figure*}

We measure $\delta /2\pi$ using transmission spectroscopy of the cavity at fixed $\Delta_{D2\pi}$. To this end we send a probe beam matched to the TEM$_{00}$ mode of the cavity, sweep its frequency by $20$ \si{\mega\hertz} within $2$ \si{\milli\second} and record the transmitted photons on a single photon counting module.
Typical raw results are shown in figure \textbf{\ref{fig:shift}} for three different choices of the atom-cavity detuning $\Delta_{D2\pi}$. In this example, we used spin imbalanced clouds comprising $2.2 \cdot 10^5$ and $1.35 \cdot 10^5$ in $\ket{\uparrow}$ and $\ket{\downarrow}$ spin components. We use a Lorentzian fit to determine the most likely value of the dispersive shift for a given histogram. We obtain $\delta /2\pi=5.3,\,3.6$ and $1.8$ \si{\mega\hertz} for $\Delta_{D2\pi}/2\pi=6.55,\,11.55$ and $21.55$ \si{\mega\hertz} respectively. The decrease of transmitted amplitude and increase in width reflects the growing weight of the atomic absorption compared with the transmission through the mirrors as the detuning is reduced, as expected in our narrow cavity regime. 

\subsection{Weakly destructive and time-resolved measurement of atomic population}

\begin{figure*}[ht]
\begin{center}
    \includegraphics{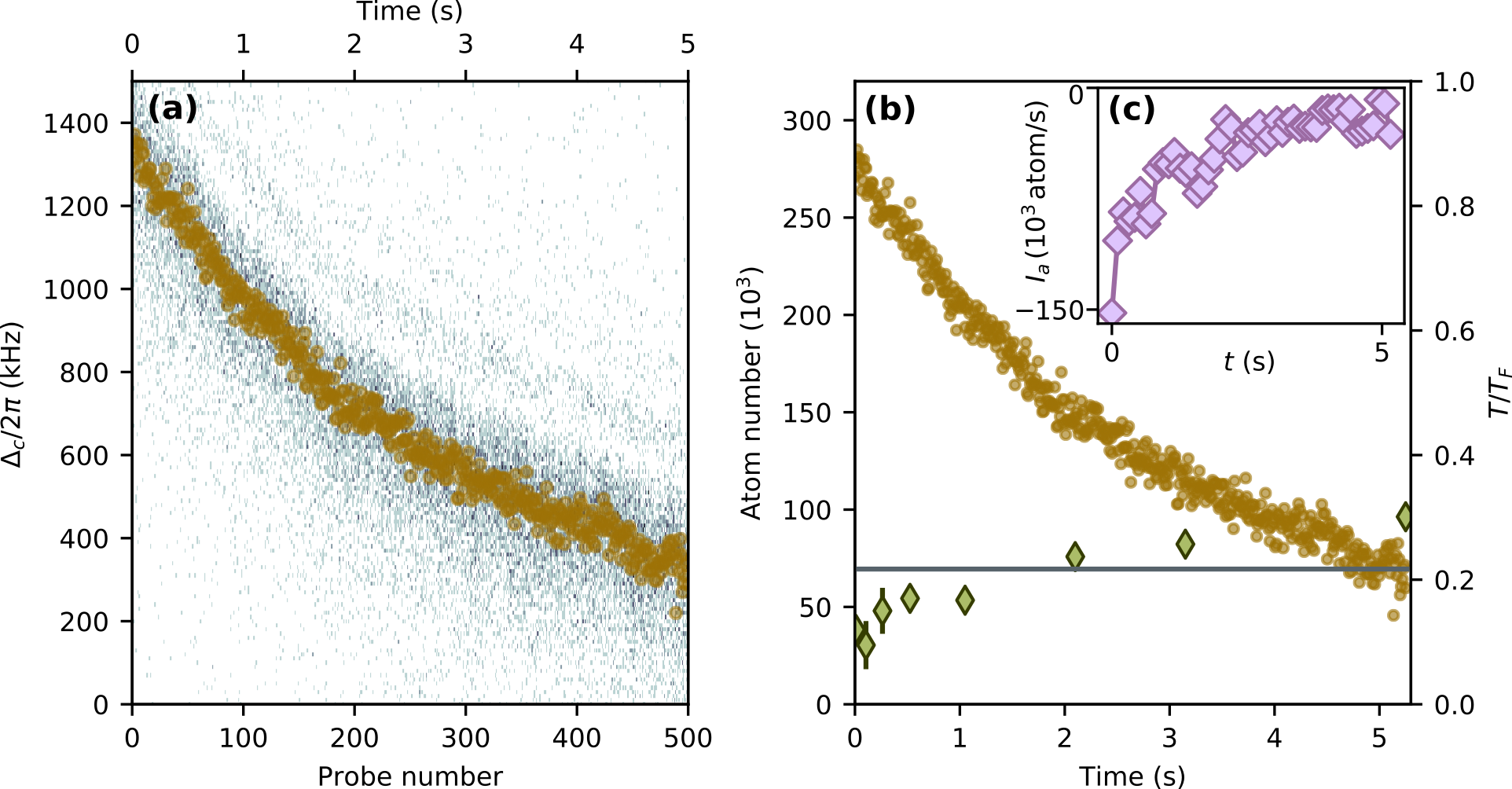}
    \caption[Destructivity]{\textbf{Repeated dispersive measurement and destructivity.}
     \textbf{(a)} Stacked raw histograms of photon counts for $500$ consecutive dispersive measurements realized on the same cloud with $\Delta_a/2\pi = 20$  \si{\giga\hertz}. The gas comprises $1.7(1) \cdot 10^5$ and $0.9(1) \cdot 10^5$ in the majority and minority spin component, respectively. Brown circles indicate the fitted location of the cavity resonance, with respect to the empty cavity resonance (see text for details).
\textbf{(b)} Inferred atom number (brown circles) as a function of time. Green diamonds indicate the temperature measured separately after  $1$, $10$, $25$, $50$, $100$, $200$, $300$ and $500$ consecutive measurements. The grey line indicates the critical temperature $T_c = 0.217 T_F$. \textbf{(c)} Atomic current calculated from the atom number variations averaged over three clouds. Each value of the derivative is fitted over $30$ consecutive dispersive probe measurements.}
    \label{fig:multiplescans}
\end{center}
\end{figure*}

The measurement of dispersive shift probes atom number while limiting resonant light scattering: as the cooperativitiy of the cavity is larger than one, a majority of the light is scattered into the cavity mode, contributing coherently to the measurement signal, as opposed to scattering into free space which amounts to incoherent losses. In addition, this technique is free of saturation and Doppler effects that hinder absorption imaging for light species \cite{Horikoshi:2017ab}. 

As we now show, this cavity-based detection offers the opportunity to monitor atom number variations on one single quantum degenerate Fermi gas over time. In order to later infer temperature increases, we use spin imbalanced gases comprising $1.7(1) \cdot 10^5$ and $9(1) \cdot 10^4$ atoms in $\ket{\uparrow}$ and $\ket{\downarrow}$ spin states. We then repeatedly send light on the cavity and sweep its frequency accross the cavity resonance, following the protocol described above, and record the transmitted signal. We performed up to $500$ measurements in total separated by $10$ \si{\milli\second}. Figure\textbf{ \ref{fig:multiplescans}(a)} presents the raw photon detection histograms obtained for all the successive sweeps over one single realization of the Fermi gas, for $\Delta_{D2\pi} = 20$ \si{\giga\hertz}. The presence of a clear resonance is obvious for each scan, with noise originating from the finite photon count. A weak ringing can be observed due to the sweep rate being comparable with the cavity linewidth. For each sweep, we determine the most likely location of the resonance using a Lorentzian fit, shown with brown circles in figure \textbf{\ref{fig:multiplescans}}a. 

These measurements directly translate into variations of atom number, as shown in figure \textbf{\ref{fig:multiplescans}(b)}. This slow decay of atom number is primarily due to the combined effects of the dipole trap spontaneous emission and intensity noise, and background gas collisions. We also repeated these measurements, stopping after a variable number of probe pulses to evaluate the heating, using the method described above. After a single scan we observe no detectable atom losses and a temperature of $T/T_F = 0.09(1)$ compatible with the one measured without any cavity probe. After $100$ measurements we measure a temperature increase to $T/T_F = 0.16(1)$, still below the superfluid critical temperature.

The losses of atoms in time represents a particle current escaping the trap, such that this measurement can be interpreted as probing atomic currents. To substantiate this, we use a linear fit of the atom number evolution over $30$ consecutive shots to extract the total current. The result is shown in figure \textbf{\ref{fig:multiplescans}(c)}, averaged over three realizations of the gas. 

To assess the role of measurements in the loss processes, we performed similar measurements with a reduced probe rate but keeping the total observation duration at $5$ \si{\second}. We observed that increasing the number of measurements from $10$ to $500$ leads to an increase by $7\%$ of the observed atomic losses (see Appendic C). Comparing the losses with different probe numbers, we estimate that a single probe pulse induces a loss of about $30$ atoms. By comparison, the standard deviation in the determination of the most likely population in the cloud represents about $3000$ atoms for the data of figure \textbf{\ref{fig:multiplescans}}. Depending on the requirements, future experiments may use larger probe power at the cost of an increased destructivity.


\section{Conclusion}
 
We have presented an apparatus combining unitary Fermi gases with a high finesse optical cavity. Using the cavity as a deep dipole trap circumvents the need of very high power lasers, allowing for the fast production of large unitary Fermi gases with reduced laser power. Our lattice cancellation scheme opens the way towards the use of cavity-enhanced traps all the way to quantum degeneracy, which would further simplify experimental schemes and further reduce the laser power requirements and thus the costs. Even in a single chamber design, the cavity does not restrict the available optical access, such that the addition of a high-resolution microscope to the setup can be envisioned without significant technical changes. 

We also demonstrated the ability to perform weakly destructive atom counting in Fermi gases. A very interesting perspective is the study of noise, in particular originating from quantum fluctuations of the atomic density overlap with the mode function, and the dynamical and possibly quantum measurement back-action. Technically, the measurements demonstrated in this paper could be used to further stabilize atom number during evaporation as demonstrated for Bosons \cite{Gajdacz:2016aa}, monitor various non equilibrium processes \cite{Wigley:2016aa,Sawyer:2017aa} or slow particle transport in the two-terminal configuration \cite{uchino_universal_2018}. The dispersive coupling of the atoms to the cavity could also be used to probe transport in a lattice by coupling to the weak density variations induced by changes of quasi-momentum \cite{Kesler:2016aa}, or directly to current via cavity-photon assisted tunneling \cite{laflamme_continuous_2017}.

Beyond these direct applications, a quantum degenerate Fermi gas dispersively coupled to the cavity field opens fascinating perspectives for the engineering of novel many-body phases. Superradiant phases with zero threshold have been predicted at commensurability \cite{keeling_fermionic_2014,chen_superradiance_2014,piazza_umklapp_2014}, with important effects of atomic interactions \cite{Chen:2015aa}. Various types of magnetically and density ordered or superfluid phases have been predicted \cite{colella_quantum_2018,colella_antiferromagnetic_2019,sheikhan_cavity-induced_2019,schlawin_cavity-mediated_2019-1}, as well as emergent chiral currents and topological features \cite{Pan:2015aa,kollath_ultracold_2016,Sheikhan:2016aa,Zheng:2016aa,mivehvar_superradiant_2017,Colella:2019ab}. Beyond these examples, the combination of a strongly correlated superfuid with the long range cavity-mediated interactions will provide an ideal platform to study competing orders in Fermionic systems. 

\ack

We acknowledge Barbara Cilenti for help at the beginning of the project, Tobias Donner and Tilman Esslinger for discussions. We acknowledge funding from the European Research Council (ERC) under the European Union’s Horizon 2020 research and innovation programme (grant agreement No 714309), the Swiss National Science Foundation (grant No 184654), the Sandoz Family Foundation-Monique de Meuron program for Academic Promotion and EPFL.

\appendix

\section*{Appendix A : Laser system}

\begin{figure}[h]
\begin{center}
    \includegraphics{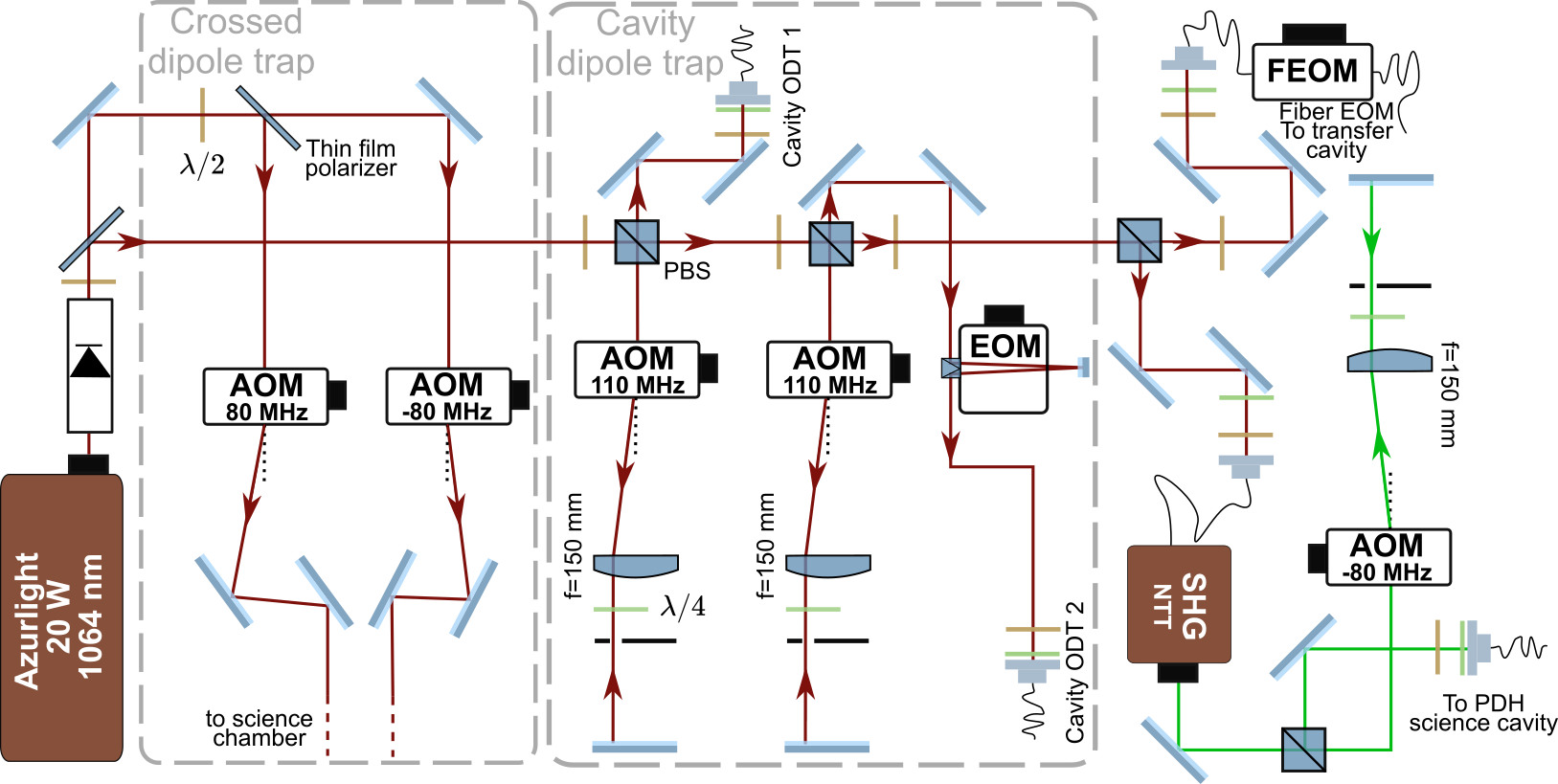}
    \caption[LaserSystem_1064]{\textbf{High-power $1064$ nm laser setup.}}
    \label{fig:LaserSystem_1064}
\end{center}
\end{figure}

Figure \textbf{\ref{fig:LaserSystem_1064}} presents the details of the 1064 nm laser system, distributed between the two arms of the crossed optical dipole trap, the two cavity dipole traps (with and without lattice cancellation), and the cavity stabilization system. Our fiber amplifier is capable of delivering 20 W but we run it at lower power, as decribed in the text. For the lattice-free cavity trap dipole trap, a free space EOM is used in double pass configuration, in order to achieve the high modulation depth needed to cancel the lattice structure without the limitation of laser power inherent to fibered EOMs.

\begin{figure}[h]
\begin{center}
    \includegraphics{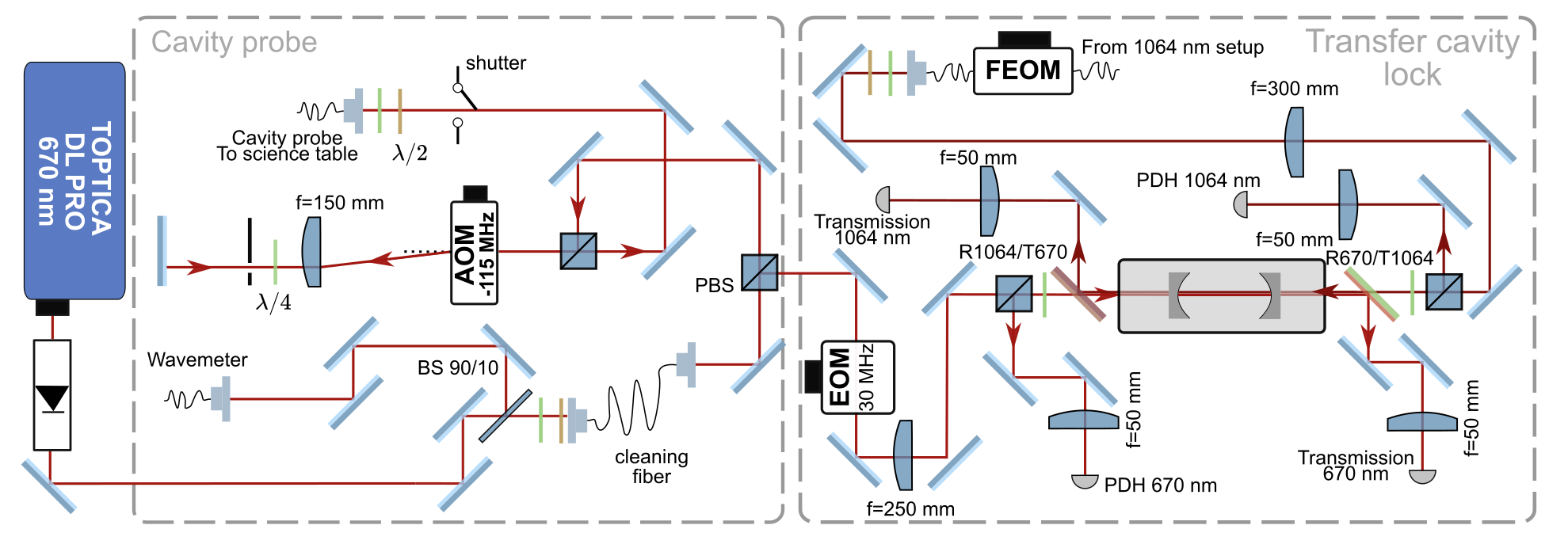}
    \caption[LaserSystem_transfercavity]{\textbf{Transfer cavity laser setup for optical dipole trap and cavity stabilization.}}
    \label{fig:LaserSystem_transfercavity}
\end{center}
\end{figure}

Figure \textbf{\ref{fig:LaserSystem_transfercavity}} shows the details of the probe laser and cavity stabilization scheme. An AOM directly after the laser is used to stabilize the power against long term drifts, so that we get shot-noise limited reproducibility for the short probe pulses. The transfer cavity itself is home made, based on mirrors similar to that of the science cavity, with a longer separation yielding a $30$ \si{\kilo\hertz} linewidth. Its length is stabilized on a sideband of the $1064$ \si{\nano\meter} laser generated by the fibered EOM. The modulation is mixed with another RF signal, allowing to generate the PDH error signal after demodulation. The probe laser is stabilized to the transfer cavity using an ultra-fast feedback controller (Toptical FALC). 

\section*{Appendix B : Thermal effect of intracavity dipole trap}

As a result of large intracavity intensity and large radius of curvature, we observed non linear effects on the cavity transmission at $1064$ \si{\nano\meter} as a function of the incoming beam frequency.
 
\begin{figure}[h]
\begin{center}
    \includegraphics{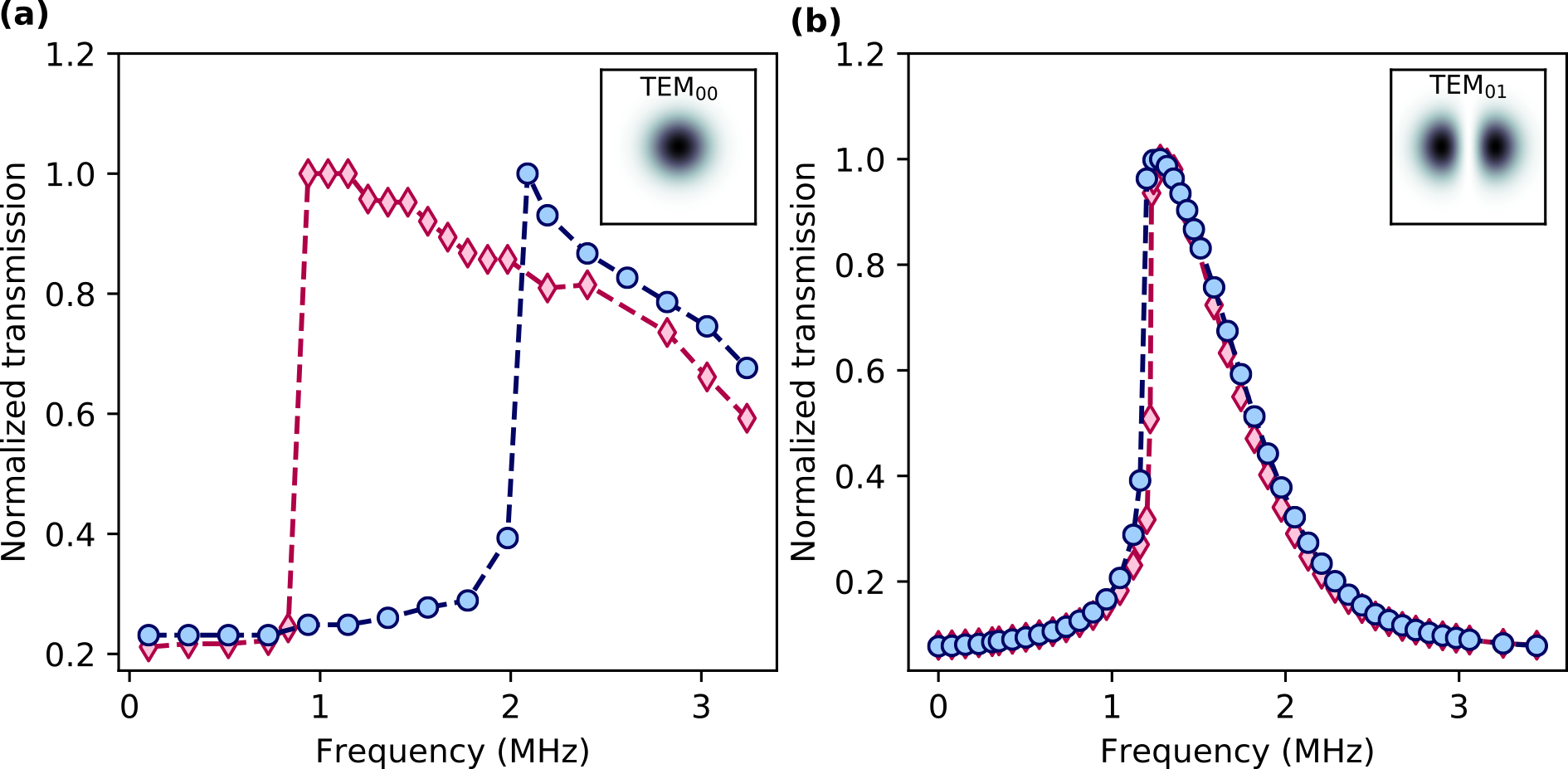}
    \caption[Thermal effects]{\textbf{Thermal effect in mirror substrates for different cavity tranverse mode.}
    We show the variation of the power for a beam coupled to the cavity mode TEM$_{00}$ \textbf{(a)} and TEM$_{01}$ \textbf{(b)}. In both cases incident power on the cavity is set to $219$ \si{\milli\watt}. In each situation we scan the frequency of the dipole trap beam for increasing (blue) and decreasing frequency (pink). In the case of the TEM$_{00}$ mode a bistable and hysteretic behaviour appears because of thermal effects in the mirrors. In contrast the use of a TEM$_{01}$ mode, which has a $2$ times larger surface on the mirrors, reduces the intensity and consequently mitigate this effect. This is necessary in order to stabilize the intracavity trap power by measuring the cavity output power to prevent parametric heating. }
    \label{fig:hysteresis}
\end{center}
\end{figure}

For an injected TEM$_{00}$ mode, a strong bistable and hysteretic behaviour is observed as shown in figure \textbf{\ref{fig:hysteresis}(a)}. This represents a significant source of preparation noise because of large heating rate in the cavity dipole trap. We circumvent this problem by injecting a beam mode matched with a TEM$_{01}$. This has two effects (i) it increase by a factor $2$ the area on which the intracavity power is spread on the mirror and (ii) it increases the trap volume by also a factor $2$ at the cost of a smaller coupling efficiency ($80\%$ for the TEM$_{00}$ compared with $50\%$ for the TEM$_{01}$ case). The larger trap volume allows to achieve the same capture efficiency from the MOT but at a lower total power and with a smaller intensity on the mirror. Both effects combined we obtain a larger cloud with negligible hysteretic effect as shown in figure \textbf{\ref{fig:hysteresis}(b)}.

This technique could be a way to allow for smaller cavity mode volume, increasing the light-matter coupling strength, with the possibility of intracavity optical trap with large mode volume taking advantage of higher-order transverse modes.

\section*{Appendix C : Effect of probing repetition rate on the destructivity of cavity-based measurement}

\begin{figure}[h]
\begin{center}
    \includegraphics{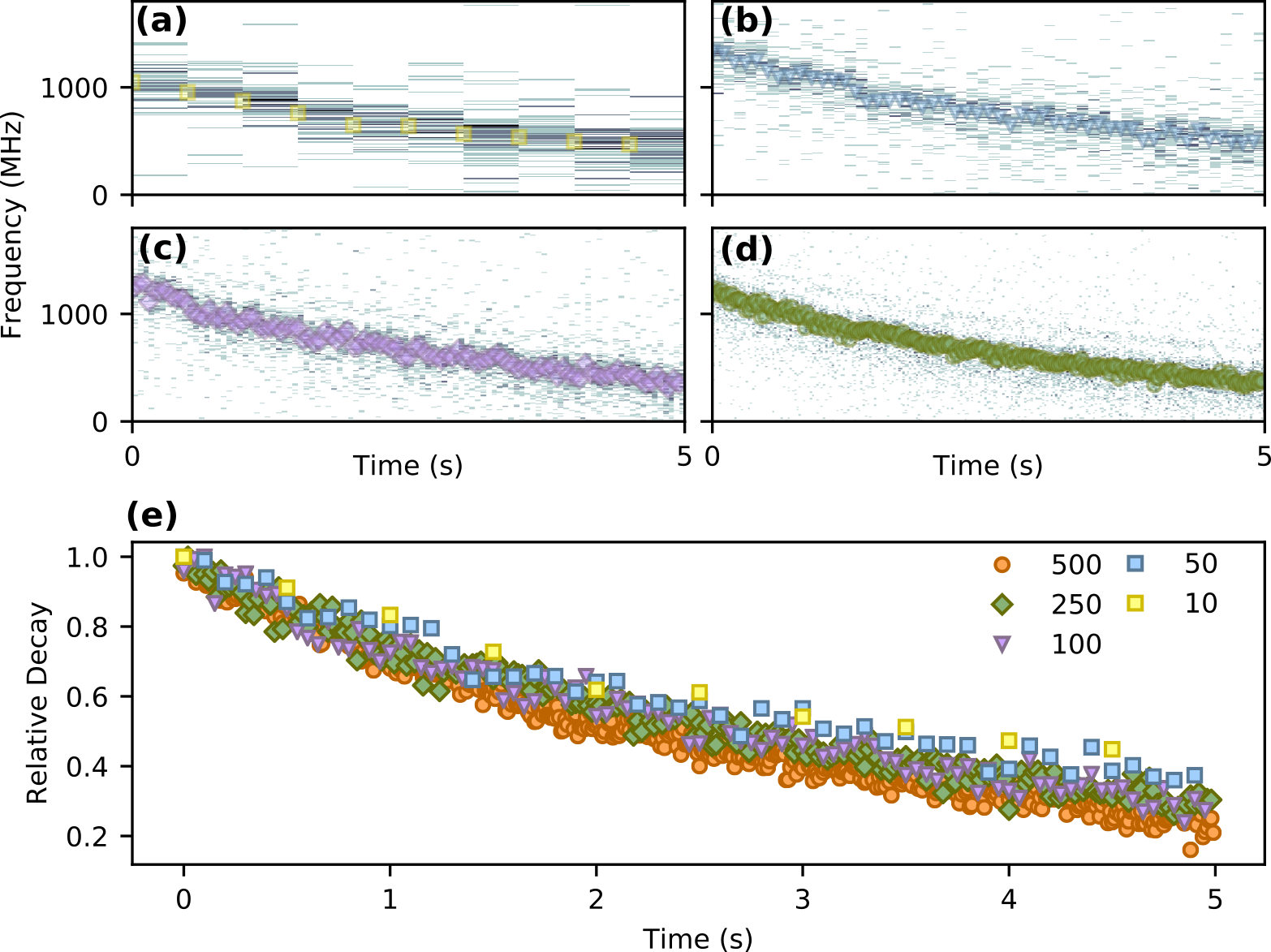}
    \caption[Destructivity]{\textbf{Effect of multiple consecutive probes on the time evolution of atom number.}
 \textbf{(a)-(d)} Evolution of the dispersive shift position over $5$ \si{\second} for respectively $10$, $50$, $100$ and $250$ consecutive measurements. Transmission spectra are aligned in time for different number of consecutive probe sweeps. The markers are the fitted position of the transmission peak using a lorentzian fit on the spectrums with a binning of $10$ \si{\kilo\hertz} and an average photon count per sweep of $55$. \textbf{(e)} Normalized time evolution of the transmission peak position for $10$, $50$, $100$, $250$ and $500$ consecutive measurement. We observe a small additional decay of peak position of $7\%$ in the case of $500$ measurement compared with the other cases.
}
    \label{fig:Destructivity}
\end{center}
\end{figure}

We study the destructivity of the cavity probe by measuring the dispersive shift at $\Delta_a /2\pi = 20$ \si{\giga\hertz} for various number of consecutive sweeps. The interval between the probe sweeps is adapted to keep the total duration of $5$ \si{\second}. We record the transmission peak positions for different number of consecutive probes as an indication of the atomic losses occuring during the measurement. Results are presented in figure \textbf{\ref{fig:Destructivity}}.

We present the time evolution of the transmission peak positions normalized by their initial value in figure \textbf{\ref{fig:Destructivity}(e)}. By comparing the relative shifts, we can deduce the destructivity of a single measurement on the atomic cloud. After $4.5$ \si{\second} the dispersive shift with $500$ consecutive probe sweeps decays $7\%$ more than with $10$ sweeps. This is a good indication of a weakly destructive measurement and compatible with an additional loss of $30$ atoms per probe sweep.

\section*{References}

\bibliographystyle{iopart-num}
\bibliography{Paper}

\end{document}